\documentclass[preprint,aps,10pt]{revtex4}
\usepackage{amssymb}
\usepackage{amsmath}
\usepackage{amsfonts}
\usepackage{graphicx}
\usepackage{graphics}
\usepackage{psfig}

\begin{document}

\title{Non-Minimal Coupling to a Lorentz-Violating Background and
Topological Implications}
\author{H. Belich}
\email{belich@cce.ufes.br}
\affiliation{{\small {Universidade Federal do Esp\'{\i}rito Santo (UFES), Departamento de
F\'{\i}sica e Qu\'{\i}mica, Av. Fernando Ferrari, S/N Goiabeiras, Vit\'{o}%
ria - ES, 29060-900 - Brasil}}}
\affiliation{{\small {Grupo de F\'{\i}sica Te\'orica Jos\'e Leite Lopes, C.P. 91933, CEP
25685-970, Petr\'opolis, RJ, Brasil}}}

\author{T. Costa-Soares}
\email{tcsoares@cbpf.br}
\affiliation{{\small {~CCP, CBPF, Rua Xavier Sigaud, 150, CEP 22290-180, Rio de Janeiro,
RJ, Brasil}}}
\affiliation{{\small {Universidade Federal de Juiz de Fora (UFJF), Col\'{e}gio T\'{e}%
cnico Universit\'{a}rio, av. Bernardo Mascarenhas, 1283, Bairro F\'{a}brica
- Juiz de Fora - MG, 36080-001 - Brasil}}}
\affiliation{{\small {Grupo de F\'{\i}sica Te\'orica Jos\'e Leite Lopes, C.P. 91933, CEP
25685-970, Petr\'opolis, RJ, Brasil}}}

\author{M.M. Ferreira Jr.}
\email{manojr@ufma.br}
\affiliation{{\small {Universidade Federal do Maranh\~{a}o (UFMA), Departamento de F\'{\i}%
sica, Campus Universit\'{a}rio do Bacanga, S\~{a}o Luiz - MA, 65085-580 -
Brasil}}}
\affiliation{{\small {Grupo de F\'{\i}sica Te\'orica Jos\'e Leite Lopes, C.P. 91933, CEP
25685-970, Petr\'opolis, RJ, Brasil}}}
\author{J.A. Helay\"{e}l-Neto}
\email{helayel@cbpf.br}
\affiliation{{\small {~CCP, CBPF, Rua Xavier Sigaud, 150, CEP 22290-180, Rio de Janeiro,
RJ, Brasil}}}
\affiliation{{\small {Grupo de F\'{\i}sica Te\'orica Jos\'e Leite Lopes, C.P. 91933, CEP
25685-970, Petr\'opolis, RJ, Brasil}}}

\begin{abstract}
The non-minimal coupling of fermions to a background responsible for the
breaking of Lorentz symmetry is introduced in Dirac's equation; the
non-relativistic regime is contemplated, and the Pauli's equation is used to
show how an Aharonov-Casher phase may appear as a natural consequence of the
Lorentz violation, once the particle is placed in a region where there is an
electric field. Different ways of implementing the Lorentz breaking are
presented and, in each case, we show how to relate the Aharonov-Casher phase
to the particular components of the background vector or tensor that
realises the violation of Lorentz symmetry.
\end{abstract}

\maketitle

\section{Introduction\-}

In the beginning of the\ nineties, Carroll-Field-Jackiw \cite{Jackiw} have
considered a Chern-Simons-like odd-CTP term able to induce the violation of
Lorentz symmetry in $(1+3)$\ dimensions. In a more recent context, some
authors \cite{Kostelec1}\ have explored the possibility of Lorentz-symmetry
breaking in connection with string theories. Models with Lorentz- and
CPT-breakings were also used as a low-energy limit of an extension of the
Standard Model, valid at the Plank scale \cite{Colladay}. In this case, an
effective action is obtained that incorporates CPT and Lorentz violation and
keeps unaffected the $SU(3)\times SU(2)\times U(1)$ gauge structure of the
underlying theory. This fact is of relevance in the sense it indicates that
the effective model may preserve some properties of the original theory,
like causality and stability. In spite of Lorentz symmetry is closely
connected to stability and causality in modern field theories, the existence
of a causal and unitary model with violation of Lorentz symmetry is in
principle possible and meaningful on physical grounds.

In the latest years, Lorentz-violating theories have been investigated under
diverse perspectives. In the context of $N=1$ - supersymmetric models, there
have appeared two proposals: one which violates the algebra of supersymmetry
(first addressed by Berger \& Kostelecky \cite{Berger}), and other that
preserves the (SUSY) algebra and yields the Carroll-Field-Jackiw model by
integrating on Grassmann variables \cite{Nos}. The study of radiative
corrections arising from the axial coupling of charged fermions to a
constant vector has revealed a controversy on the possible generation of the
Chern-Simons-like term that has motivated a great deal of works in the
literature \cite{Chung}. The rich phenomenology of fundamental particles has
also been considered as a natural environment to the search for indications
of breaking of these symmetries \cite{Coleman}, \cite{Particles}, revealing
up the moment stringent limitations on the factors associated with such
violation. The traditional discussion concerning space-time varying coupling
constants has also been addressed to in the light of Lorentz-violating
theories\cite{Coupling}, with interesting connections with the construction
of Supergravity. Moreover, measurements of radio emission from distant
galaxies and quasars put in evidence that the polarizations vectors of the
radiation emitted are not randomly oriented as naturally expected. This
peculiar phenomenon suggests that the space-time intervening between the
source and observer may be exhibiting some sort of optical activity
(birefringence), whose origin is unknown \cite{Astrophys}. Some different
proposals for implementation of Lorentz violation have been recently
considered; one of them consists in obtaining this breaking from spontaneous
symmetry breaking of a vector matter field\ \cite{colatto}.

Our approach to the Lorentz breaking consists in adopting the 4-dimensional
version of a Chern-Simons topological term, namely $\epsilon _{\mu \nu
\kappa \lambda }v^{\mu }A^{\nu }F^{\kappa \lambda }$ (also known as the
Carroll-Field-Jackiw\ term \cite{Jackiw}), where $\epsilon _{\mu \nu \kappa
\lambda }$ is the 4-dimensional Levi-Civita symbol and $v^{\mu }$ is a fixed
four-vector acting as a background. A study of the consequences of such
breaking in a QED has been extensively analyzed in literature \cite{Adam}, 
\cite{Chung}. An extension of the Carroll-Field-Jackiw (CFJ)\ model to
include a scalar sector that yields spontaneous symmetry breaking (Higgs
sector) was recently developed and analyzed, resulting in an Abelian-Higgs
CFJ\ electrodynamics (AHCFJ model) with violation of Lorentz symmetry \cite%
{Belich}. Afterwards, the dimensional reduction of the CFJ\ and AHCFJ models
(to 1+2 dimensions) were successfully carried out in refs. \cite{Manojr1}, 
\cite{Manojr3},\ respectively, resulting in a planar Maxwell-Chern-Simons
and Maxwell-Chern-Simons-Proca electrodynamics coupled to a scalar field by
a mixing term (responsible for the Lorentz violation). It should be here
stressed that these planar models do not present the causality and unitarity
problems that affect the original CFJ and AHCFJ models; instead they
constitute entirely consistent planar theories, whose properties have been
recently investigated under different aspects \cite{Belich2}, \cite{Manojr4}.

Topological effects in quantum mechanics are phenomena that present no
classical counterparts, being associated with physical systems defined on a
multiply connected space-time. Especifically, considering a charged particle
that propagaties in a region with external magnetic field (free force
region),\ it is verified that the corresponding wave function may develop a
quantum phase: $\langle b|a\rangle _{inA}=\langle b|a\rangle _{A=0}.\left\{
\exp (iq\int_{a}^{b}\mathbf{A.dl}\text{ })\right\} ,$ which describes the
real behavior of the electron's propagation. This issue has received
considerable attention since the pioneering work by Aharonov and Bohm \cite%
{Bohm}, where they demonstrated that the vector potential may induce
physical measurable quantum phases even in a field-free region, which
constitute the essence of a topological effect. The induced phase does not
depend on the specific path described by the particle neither on its
velocity (nondispersiveness). Instead, it is intrinsically related to the
non-simply connected nature of the space-time and to the associated winding
number. Many years later, Aharonov and Casher \cite{Casher} argued that a
quantum phase also appears in the wave function of a neutral spin$-\frac{1}{2%
}$ particle with anomalous magnetic moment, $\mu $, subject to an electric
field arising from a charged wire. This is the well-known Aharonov-Casher
(A-C) effect, which is related to the A-B effect by a duality operation.

This effect can be obtained by taking into account the non-relativistic
limit of the Dirac's equation \cite{Anandan} with the Pauli-type non-minimal
coupling. Concerning these phase effects, other developments over the past
years have raised a number of interesting questions, in connection to which
locality and topology are being invoked in a more recent context \cite{Topol}%
. The local or topological nature of the generated phase can change
according to each situation, as in the case of the A-B effect in molecular
systems, which is neither local nor topological, being more closely to the
A-C effect. For instance, the work of ref. \cite{Prd40} discusses the A-C
phase in a \ planar model in order to demonstrate that this effect is
essentially non-local in the context of a non-relativistic superconductor.
The formal correspondence between the A-B and A-C phases at a microscopic
level, as long as their topological nature is concerned, is considered in
ref. \cite{Choi}. In the context of ultra-cold atoms, it was shown that the
vortex model of Bose-Einstein Condensates is described by a Lagrangian with
an A-C extra term \cite{Bec}. \ 

In this work, we propose the investigation of non-minimal coupling terms in
the context of Lorentz-violating models involving some fixed background and
the gauge and fermion fields. The main purpose is to figure out whether such
new couplings are able to induce the A-C effect. In this sense, we follow a
single procedure: writing the spinor field in terms of its weak and strong
components, we achieve the Pauli equation (once the non-relativistic limit
of the Dirac equation is considered) and identify the generalized canonical
momentum, which in this approach plays a central role for determination of
the induced quantum phases.

This paper is organized as follows. In the first section, several kinds of
Lorentz-violating non-minimal couplings are analyzed in connection with the
possibility of\thinspace\ generating an A-C quantum phase. Initially, we
consider the presence of the non-minimal term, $igv^{\nu }\overset{\sim }{F}%
_{\mu \nu },$ in the covariant derivative\ - reflecting the coupling of a
neutral test particle to the Lorentz breaking background. In the sequence,
taking the non-relativistic limit, we derive the Pauli equation and write
the generalized canonical momentum, whose composition implies the appearance
of an A-C phase. \ In this case, the background 3-vector plays the role of
the magnetic moment of the neutral particle. As a second case, \ we regard a
non-minimally torsion-like ($\gamma _{5}$-type) coupling with the
Lorentz-violating background in the context of the Dirac equation. No A-C
phase is generated in this case. In another situation, a background tensor ($%
T_{\mu \nu })$ responsible for the violation of Lorentz symmetry is
non-minimally coupled to the electromagnetic and Dirac fields. It is
observed that the anti-symmetric part of this tensor is associated with the
induction of an A-C phase. As a final investigation, it was simultaneously
considered the non-minimal Lorentz-breaking coupling and the (Pauli)
standard non-minimal coupling in the context of the Dirac equation in order
to study the competition among both these terms in connection with the
generation of an A-C phase. It has been then verified that only the standard
Pauli magnetic coupling yields an A-C effect. \ Our Final Discussions are
cast in Section 3.

\section{Lorentz-violating on-minimal couplings, Pauli equation and the
Aharonov-Casher phase}

\subsection{Non-minimal coupling to the gauge field and background}

The first case to be analyzed starts with the gauge invariant Dirac
equation, 
\begin{equation}
(i\gamma ^{\mu }D_{\mu }-m)\Psi =0,  \label{Dirac1}
\end{equation}%
where the covariant derivative with non-minimal coupling is chosen to be 
\begin{equation}
D_{\mu }=\partial _{\mu }+eA_{\mu }+igv^{\nu }\overset{\sim }{F}_{\mu \nu },
\label{covader}
\end{equation}%
whereas $v^{\mu }$\ is a fixed four-vector acting as the background which
breaks the Lorentz symmetry\cite{Jackiw}. The explicit representation of the 
$\gamma $-matrices used through out is listed below:

\begin{equation}
\gamma ^{0}=\left( 
\begin{tabular}{ll}
$1$ & $0$ \\ 
$0$ & -$1$%
\end{tabular}%
\ \ \right) ,\overset{\rightarrow }{\gamma }=\left( 
\begin{tabular}{ll}
$0$ & $\overset{\rightarrow }{\sigma }$ \\ 
-$\overset{\rightarrow }{\sigma }$ & $0$%
\end{tabular}%
\ \ \right) ,\gamma _{5}=\left( 
\begin{tabular}{ll}
$0$ & $1$ \\ 
$1$ & $0$%
\end{tabular}%
\ \ \right) ,
\end{equation}%
where $\overrightarrow{\sigma }=(\sigma _{x},\sigma _{y},\sigma _{z})$ are
the Pauli matrices. \ In order to simplify the calculations, the
spinor\thinspace $\Psi $\ should be written in terms of small $\left( \chi
\right) $\ and large $\left( \phi \right) $\ two-spinors, $\Psi =\left( 
\begin{tabular}{l}
$\phi $ \\ 
$\chi $%
\end{tabular}%
\ \ \right) ,$\ so that eq. (\ref{Dirac1}) splits into two equations for $%
\phi $ and $\chi $: 
\begin{eqnarray}
\left( E-e\varphi -g\overrightarrow{v}\cdot \overrightarrow{B}\right) \phi -%
\overrightarrow{\sigma }\cdot (\overrightarrow{p}-e\overrightarrow{A}+gv^{0}%
\overrightarrow{B}-g\overrightarrow{v}\times \overrightarrow{E})\chi 
&=&m\phi ,  \label{phi1} \\
-\left( E-e\varphi -g\overrightarrow{v}\cdot \overrightarrow{B}\right) \chi -%
\overrightarrow{\sigma }\cdot (\overrightarrow{p}-e\overrightarrow{A}+gv^{0}%
\overrightarrow{B}-g\overrightarrow{v}\times \overrightarrow{E})\phi 
&=&m\chi ,  \label{phi2}
\end{eqnarray}

\bigskip Writing the weak component in terms of the strong one (in the
non-relativistic limit), one has: 
\begin{equation}
\chi =\frac{1}{2m}\overrightarrow{\sigma }\cdot \left( \overrightarrow{p}-e%
\overrightarrow{A}+gv^{0}\overrightarrow{B}-g\overrightarrow{v}\times 
\overrightarrow{E}\right) \phi .
\end{equation}%
Replacing such a relation in eq. (\ref{phi2}), one achieves the associated
Pauli equation for the strong component $\phi $, namely: 
\begin{equation}
\left( E-e\varphi -g\overrightarrow{v}\cdot \overrightarrow{B}\right) \phi -%
\frac{1}{2m}\left( \overrightarrow{\sigma }\cdot \overrightarrow{\Pi }%
\right) \left( \overrightarrow{\sigma }\cdot \overrightarrow{\Pi }\right)
\phi =m\phi ,  \label{Pauli1}
\end{equation}%
where the canonical generalized moment is defined as, 
\begin{equation}
\overrightarrow{\Pi }=\left( \overrightarrow{p}-e\overrightarrow{A}+gv^{0}%
\overrightarrow{B}-g\overrightarrow{v}\times \overrightarrow{E}\right) .
\end{equation}%
The presence of the term, $g\overrightarrow{v}\times \overrightarrow{E},$\ \
which possesses a non-null rotational, is the factor that determines the
induction of the\textbf{\ }Aharonov-Casher effect. Indeed, the 3-vector
background plays the role of a sort of magnetic dipole moment, $\left( 
\overrightarrow{\mu }=g\overrightarrow{v}\right) $, that gives rise to the
A-C phase associated with the wave function of a neutral test particle ($e=0$%
), for which the Aharonov-Bohm effect is absent. In the case of a charged
particle, the non-minimal coupling of eq. (\ref{covader}) brings about
simultaneously the A-B and A-C phases. For a neutral particle under the
action of an external electric field, the A-C phase induced as a consequence
of the Lorentz symmetry violation read as 
$\Phi _{AC} = \int ( g \overrightarrow{v} \times \overrightarrow{E}) \cdot \overrightarrow{dl}$
where c is a closed path. (eu nao sei fazer a intgral de contorno fechado no
Latex)

With this result, we can comment on another possible non-minimal coupling,
which has not been included in the covariant derivative (\ref{covader}), $%
ihv^{\nu }F_{\mu \nu }$, with $h$ being the coupling constant. It does not
yield an A-C phase, but it rather implies an extra phase involving the
magnetic field and it takes the form $\overrightarrow{v}\times 
\overrightarrow{B}$.

To write the Hamiltonian associated with the Pauli equation exhibited above,
one should use the well-known identity, 
\begin{equation}
\left( \overrightarrow{\sigma }\cdot \overrightarrow{\Pi }\right) ^{2}=%
\overrightarrow{\Pi }^{2}+i\overrightarrow{\sigma }\cdot \left( 
\overrightarrow{\Pi }\times \overrightarrow{\Pi }\right) ,  \label{identity}
\end{equation}%
which after some algebraic manipulations leads to: 
\begin{equation}
H=\frac{1}{2m}\overrightarrow{\Pi }^{2}+e\varphi -\frac{e}{2m}%
\overrightarrow{\sigma }\cdot (\overrightarrow{\nabla }\times 
\overrightarrow{A})+\frac{1}{2m}gv^{0}\overrightarrow{\sigma }\cdot (%
\overrightarrow{\nabla }\times \overrightarrow{B})+\frac{g}{2m}%
\overrightarrow{\sigma }\cdot \overrightarrow{\nabla }\times (%
\overrightarrow{v}\times \overrightarrow{E}).
\end{equation}

\subsection{Torsion Non-Minimal Coupling with Lorentz\ Violation}

In this section, one deals again with eq. (\ref{Dirac1}), now considering
another kind of non-minimal coupling, 
\begin{equation}
D_{\mu }=\partial _{\mu }+eA_{\mu }+ig_{a}\gamma _{5}v^{\nu }\overset{\sim }{%
F}_{\mu \nu },
\end{equation}%
where the Lorentz-violating background, $v^{\mu }$, appears coupled to the
gauge field by means of a torsion-like term of chiral character. \ 

Writing the spinor $\Psi $\ in terms of the so-called small and large
components in much the same way of the latter section, there appear two
coupled equations for the 2-component spinors $\phi ,\chi $, 
\begin{eqnarray}
\left[ \left( E-e\varphi \right) +\overrightarrow{\sigma }\cdot \left(
g_{a}v^{0}\overrightarrow{B}-g_{a}\overrightarrow{v}\times \overrightarrow{E}%
\right) \right] \phi -[\overrightarrow{\sigma }\cdot \left( \overrightarrow{p%
}-e\overrightarrow{A}\right) +g_{a}\overrightarrow{v}\cdot \overrightarrow{B}%
]\chi =m\phi , &&  \label{phi3} \\
-\left[ \left( E-e\varphi \right) +\overrightarrow{\sigma }\cdot \left(
g_{a}v^{0}\overrightarrow{B}-g_{a}\overrightarrow{v}\times \overrightarrow{E}%
\right) \right] \chi +[\overrightarrow{\sigma }\cdot \left( \overrightarrow{p%
}-e\overrightarrow{A}\right) +g_{a}\overrightarrow{v}\cdot \overrightarrow{B}%
]\phi  &=&m\chi ,  \label{phi4}
\end{eqnarray}%
from which we can read the weak component in terms of the strong one: 
\begin{equation}
\chi =\frac{1}{2m}\left[ \overrightarrow{\sigma }\cdot \left( 
\overrightarrow{p}-e\overrightarrow{A}\right) +g_{a}\overrightarrow{v}\cdot 
\overrightarrow{B}\right] \phi .  \label{phi5}
\end{equation}%
From eqs. (\ref{phi4},\ref{phi5}), one obtains the correlate Pauli equation, 
\begin{equation}
\left( E-e\varphi +\overrightarrow{\sigma }\cdot \left( g_{a}v^{0}%
\overrightarrow{B}-g_{a}\overrightarrow{v}\times \overrightarrow{E}\right)
\right) \phi -\left[ \overrightarrow{\sigma }\cdot \left( \overrightarrow{p}%
-e\overrightarrow{A}\right) +g_{a}\overrightarrow{v}\cdot \overrightarrow{B}%
\right] \frac{1}{2m}\left[ \overrightarrow{\sigma }\cdot \left( 
\overrightarrow{p}-e\overrightarrow{A}\right) +g_{a}\overrightarrow{v}\cdot 
\overrightarrow{B}\right] \phi =m\phi ,
\end{equation}%
whose structure reveals as canonical generalized moment the usual relation,$%
\overrightarrow{\text{ }\Pi }=\left( \overrightarrow{p}-e\overrightarrow{A}%
\right) .$ Here, one notices that the non-minimal coupling gives rise only
to an energy contribution denoted by $H_{nm}$ and given as below 
\begin{equation}
H_{nm}=\overrightarrow{\sigma }\cdot \left( g_{a}v^{0}\overrightarrow{B}%
-g_{a}\overrightarrow{v}\times \overrightarrow{E}\right) +g_{a}%
\overrightarrow{v}\cdot \overrightarrow{B}
\end{equation}%
so that the Hamiltonian becomes{\LARGE \ }$H=\frac{1}{2m}\overrightarrow{\Pi 
}^{2}+e\varphi -\frac{e}{2m}\overrightarrow{\sigma }\cdot \nabla \times 
\overrightarrow{A}+H_{nm}.$

We can thus conclude that, if the fixed background is associated with the
vector component of the torsion, as done in the work of ref. \cite{Sh-ryder}%
, no A-C phase is induced. The coupling to the torsion contributes to the
interaction energy, but contrary to the case\ contemplated in the previous
section, the $\gamma _{5}$-type non-minimal coupling \ does not bring about
any A-C phase.

\subsection{Lorentz-violating Non-Minimal Coupling to a Tensor Background}

The starting point now is an extended Dirac equation minimally coupled to
electromagnetic field, explicitly given by, 
\begin{equation}
(i\gamma ^{\mu }D_{\mu }-m+i\lambda _{1}T_{\mu \nu }\Sigma ^{\mu \nu
}+i\lambda _{2}T_{\mu \kappa }F^{\kappa }\text{ }_{\nu }\Sigma ^{\mu \nu
})\Psi =0,
\end{equation}%
where the covariant derivative is usual one, $D_{\mu }=\partial _{\mu
}+eA_{\mu },$\ and the bilinear term, $\Sigma ^{\mu \nu }=i[\gamma ^{\mu },$ 
$\gamma ^{\nu }]/2$, is written as: 
\begin{equation*}
\Sigma ^{0i}=i\left( 
\begin{tabular}{ll}
$0$ & $\overrightarrow{\sigma }$ \\ 
$-\overrightarrow{\sigma }$ & $0$%
\end{tabular}%
\ \ \right) ,\Sigma ^{ij}=\varepsilon _{ijk}\sigma ^{k}\left( 
\begin{tabular}{ll}
$1$ & $0$ \\ 
$0$ & $1$%
\end{tabular}%
\ \ \right) .
\end{equation*}

Notice that in this case, the skew-symmetric tensor $T_{\mu \nu }$ is the
element responsible for the Lorentz violation at the level of the fermionic
coupling.\ In analogy to what occurs when fermion couplings violate Lorentz
symmetry, by means of a term of the type $b_{\mu }\overline{\Psi }\gamma
^{\mu }\gamma _{5}\Psi $ \cite{Perez-victoria}, we here propose Lorentz
violation by taking into account fermionic couplings at the form: $\overline{%
\Psi }\Sigma ^{\mu \nu }\Psi T_{\mu \nu }$ and $\overline{\Psi }\Sigma ^{\mu
\nu }\Psi F_{\mu \kappa }T^{\kappa }$ $_{\nu }.$

Following the same procedure previously adopted, we write down 2-component
equations:

\begin{eqnarray}
&& \left( E-e\varphi \right) \phi -\overset{\rightarrow }{\sigma }_{\cdot
}\left( \overrightarrow{p}-e\overrightarrow{A}\right) \chi +4i\lambda
_{1}T_{0i}\sigma ^{i}\chi +\lambda _{1}T_{ij}\varepsilon _{ijk}\sigma
^{k}\phi +2i\lambda _{2}T^{0i}F_{ij}\sigma ^{j}\chi  \notag \\
&& + T^{i0}F_{0k}\varepsilon _{ijk}\sigma ^{j}\phi +\lambda
_{2}T^{ij}F_{jk}\varepsilon _{ijk}\sigma ^{j}\phi +2i\lambda
_{2}T^{ij}F_{j0}\sigma ^{i}\chi =m\phi \\
&&  \notag \\
&& \left( E-e\varphi \right) \chi +\overset{\rightarrow }{\sigma }_{\cdot
}\left( \overrightarrow{p}-e\overrightarrow{A}\right) \phi +4i\lambda
_{1}T_{0i}\sigma ^{i}\phi +\lambda _{1}T_{ij}\varepsilon _{ijk}\sigma
^{k}\chi +2i\lambda _{2}T^{0i}F_{ij}\sigma ^{j}\phi  \notag \\
&&+T^{i0}F_{0k}\varepsilon _{ijk}\sigma ^{j}\chi +\lambda
_{2}T^{ij}F_{jk}\varepsilon _{ijk}\sigma ^{j}\chi +2i\lambda
_{2}T^{ij}F_{j0}\sigma ^{i}\phi =m\chi
\end{eqnarray}

The pair of the 2-component equations above involves both the external
electric and magnetic filds. As the A-C effect is the main subject of
interest of this investigation, we shall consider a vanishing magnetic
field, $F_{ij}=0$, which allows to achieve the following expression relating
small and large components: \ 
\begin{equation}
\chi =\frac{1}{2m}\left[ \sigma _{\cdot }\left( \overrightarrow{p}-e%
\overrightarrow{A}\right) +\lambda _{1}4iT_{oi}\sigma ^{i}+\lambda _{2}2i(%
\overrightarrow{T}\times \overrightarrow{E})_{i}\sigma ^{i}\right] \phi ,
\label{ki}
\end{equation}%
where it was used: $T^{ij}F_{j0}=(\overrightarrow{T}\times \overrightarrow{E}%
)_{i}.$ By factoring out the Pauli matrices, $\overrightarrow{\sigma }$, the
following canonical moment can be identified: 
\begin{equation}
\overrightarrow{\Pi }=\left( \overrightarrow{p}-e\overrightarrow{A}-4\lambda
_{1}\overrightarrow{T}_{1}-2\lambda _{2}\overrightarrow{T}_{2}\times 
\overrightarrow{E}\right) ,  \label{canonical}
\end{equation}%
where we have distinguished the \textquotedblleft
electric\textquotedblright\ and \ the \textquotedblleft
magnetic\textquotedblright components of the tensor, respectively defined
as: $T_{0i}=\overrightarrow{T}_{1}$, $T^{ij}=\overrightarrow{T}_{2}$.

Replacing this in eq. (\ref{ki}) , we get the following equation for the
strong spinor component: 
\begin{equation}
\left( E-e\varphi -\lambda _{1}T_{kj}\varepsilon _{ijk}\sigma ^{i}+\lambda
_{2}T^{j0}F_{0k}\varepsilon _{ijk}\sigma ^{i}\right) \phi -\frac{1}{2m}(%
\overrightarrow{\sigma }\cdot \overrightarrow{\Pi })(\sigma \cdot 
\overrightarrow{\Pi })\phi =m\phi .
\end{equation}%
Here again two kinds of quantum phases appear, one governed by $\lambda _{1}$
and the\ other by $\lambda _{2}$. However, having in mind that our purpose
is to clarify how the A-C phase can emerge, we can take $\lambda _{1}=0$.
The $\lambda _{\dot{2}}$-term, with the \textquotedblleft
component\textquotedblright\ component of $T^{\mu \nu }$,\ gives rise to the
A-C contribution. Should we take $\lambda _{1}=0$, the Lorentz-breaking term
would not impose $T_{\mu \nu }$ to be skew-symmetric. Indeed, if $T_{\mu \nu
}$ were taken to be a general tensor, the conditions for the A-C to appear
(no magnetic field but only an external electric field) would anyhow select
the anti-symmetric magnetic component, $T_{ij}=-T_{ji}$; the A-C phase is
therefore induced by the anti-symmetric piece of $T_{\mu \nu }$.

\bigskip In this case, the Hamiltonian is given as follows: 
\begin{equation}
H=\frac{1}{2m}\overrightarrow{\Pi }^{2}+e\varphi -\frac{e}{2m}%
\overrightarrow{\sigma }\cdot \overrightarrow{\nabla }\times \overrightarrow{%
A}+\frac{1}{2m}\overrightarrow{\sigma }\cdot \overrightarrow{\nabla }\times
\left( 4i\lambda _{1}\overrightarrow{T}_{1}+2i\lambda _{2}\overrightarrow{T}%
_{2}\times \overrightarrow{E}\right) .
\end{equation}

\subsection{Competition between Lorentz-preserving and Lorentz-violating
non-minimal couplings}

In this section, we would like to compare the specific non-minimal coupling
with Lorentz violation exhibited in eqs. (\ref{Dirac1},\ref{covader}) with
the standard non-minimal coupling that generates the usual Aharonov-Casher
effect, in such a way to verify how these terms are related to the
development of an A-C phase.

The gauge invariant Dirac equation from which we shall compute the Pauli
equation is, 
\begin{equation}
(i\gamma ^{\mu }D_{\mu }-m+f\Sigma ^{\mu \nu }F_{\mu \nu })\Psi =0,
\label{Dirac4}
\end{equation}%
where the covariant derivative with non-minimal coupling is the one given in
eq. (\ref{covader}). Following the same procedure already adopted, we shall
work out the non-relativistic limit of the Dirac equation. Writing the
spinor $\Psi $ in components small and large, from eq. (\ref{Dirac4}) it
results two equations: 
\begin{eqnarray}
\left[ E-e\varphi -g\overrightarrow{v}_{\bullet }\overrightarrow{B}+\left(
2f\Sigma ^{0i}F_{0i}+f\Sigma ^{ij}F_{ij}\right) \right] \phi -%
\overrightarrow{\sigma }\cdot \left( \overrightarrow{p}-e\overrightarrow{A}%
+gv^{0}\overrightarrow{B}-g\overrightarrow{v}\times \overrightarrow{E}%
\right) \chi  &=&m\phi , \\
\left[ -(E-e\varphi -g\overrightarrow{v}_{\bullet }\overrightarrow{B}%
)+\left( 2f\Sigma ^{0i}F_{0i}+f\Sigma ^{ij}F_{ij}\right) \right] \chi +%
\overrightarrow{\sigma }\cdot \left( \overrightarrow{p}-e\overrightarrow{A}%
+gv^{0}\overrightarrow{B}-g\overrightarrow{v}\times \overrightarrow{E}%
\right) \phi  &=&m\chi .
\end{eqnarray}

In the non-relativistic limit, there appears the following Pauli equation, 
\begin{equation}
\left( E-e\varphi -g\overrightarrow{v}_{\bullet }\overrightarrow{B}%
+f\varepsilon _{ijk}\sigma ^{k}F_{ij}\right) \phi -\left( \overrightarrow{%
\sigma }\cdot \overrightarrow{\mathcal{P}}\right) \frac{1}{2m}\left( 
\overrightarrow{\sigma }\cdot \overrightarrow{\mathcal{P}}\right) =m\phi ,
\end{equation}%
where: $\overrightarrow{\mathcal{P}}=\left( \overrightarrow{p}-e%
\overrightarrow{A}+gv^{0}\overrightarrow{B}-g\overrightarrow{v}\times 
\overrightarrow{E}-2if\overrightarrow{E}\right) .$

Making use of identity (\ref{identity}), we can observe that only the $f$
coupling contributes to the canonical conjugated momentum, given as: \ 
\begin{equation}
\overrightarrow{\Pi }=\left( \overrightarrow{p}-\overrightarrow{\mu }\times 
\overrightarrow{E}\right) .
\end{equation}%
As a consequence, it is observed that only the standard Pauli coupling
contributes to the A-C phase. The Lorentz breaking non-minimal coupling in
the covariant derivative contributes here only with an extra energy term, in
the form: $4fg\overrightarrow{\sigma }\cdot \left( \left( \overrightarrow{v}%
\times \overrightarrow{E}\right) \times \overrightarrow{E}\right) $; no
phase effect is induced by the Lorentz-violating background vector.

\section{\protect\bigskip Final Discussions}

In this paper, we have carried out an analysis of the role of possible
Lorentz-violating couplings in connection with the Aharonov-Casher phase
developed by an electrically neutral particle. Usually, this phase is
induced on a neutral particle endowed with a non-trivial magnetic moment
interacting with an external electric field generated by an axial charge
distribution, but may also arise in other theoretical contexts. Indeed, it
has been here argued that in the case of a non-minimal coupling to the fixed
background $v^{\mu }$ (responsible for the Lorentz breaking), an A-C phase
is developed even by neutral spinless particles, stemming from the term $g%
\overrightarrow{v}\times \overrightarrow{E}$ (present in the canonical
momentum),\ where $g\overrightarrow{v}$\ plays the role of the intrinsic
magnetic moment of the test particle. This is in close analogy to a similar
result in $\left( 1+2\right) $-dimensional Electrodynamics: charged scalars,
non-minimally coupled to an electromagnetic field, acquire a magnetic dipole
moment \cite{Paul-khare}. In our case, the situation is more drastic: a
neutral and spinless particle acquires a magnetic moment, $g\overrightarrow{v%
}$, as a by-product of the non-minimal Lorentz-violating coupling. Other
possibilities have been taken into account as well, such as the non-minimal
coupling to the torsion tensor;\ in this case no A-C phase comes out;\
instead, we get \ an extra energy contribution due to the coupling of the
spin to the Lorentz-violating background and the electric and magnetic
fields.\ Moreover,\ for Lorentz violation at the level of the fermionic
couplings, parametrized by a skew-symmetric tensor, $T_{\mu \nu }$, it was
verified that such a coupling may yield an A-C phase if the
\textquotedblleft magnetic\textquotedblright\ component of $T_{\mu \nu }$, $%
T^{ij}=\overrightarrow{T}_{2}$ is non-vanishing. The phase generated here is
not obviously shared by scalar particles, as the kind of non-minimal
coupling leading to the phase is specific for spin-$\frac{1}{2}$\ particles.
Actually, the only non-minimal coupling universal for all types of
particles, regardless their spin, is the one given in the covariant
derivative according to eq.\ref{covader}. Finally, a remarkable result is
the competition between two non-minimal couplings which separately yield the
A-C phase. The case investigated involves the non-minimal standard Pauli
coupling and the non-minimal coupling to $v^{\mu }$ analysed in the first
section. Once both interactions are switched on, it is then observed that
the A-C phase that survives is the usual one: the one stemming from the term 
$\overrightarrow{\mu }\times \overrightarrow{E}$, where $\overrightarrow{\mu 
}$ is the canonical magnetic moment of the spin-$\frac{1}{2}$ particle.

So, as a general outcome, we can state that an interesting\ effect of
breaking Lorentz and CPT symmetries is the possibility to have direct
consequences on the A-C phase for test particles once the latter couple
non-minimally to the vector or tensor background that accomplishes the
breaking. This is a feature of Lorentz-violating gauge models not yet
discussed in the literature. We then argue that in this scenario, even
neutral scalar particles may acquire a non-trivial A-C phase once acted upon
by an external electric field, and we attribute to the Lorentz-violating
background vector, non-minimally coupled to the specific test particle, the
property of inducing the magnetic dipole moment that couples to the electric
field to give rise to the A-C phase. This result is very similar to a
mechanism that takes place in planar gauge theories. Indeed, in (1+2)D, a
number of works \cite{nogue} have shown how a scalar particle may acquire a
non-trivial magnetic moment at the expenses of a non-minimal coupling to the
Maxwell field. We can actually compare our present result to the
(1+2)-dimensional counterpart if the violation of the Lorentz symmetry takes
place due to an external 4-vector background. The latter sets up,
effectively, a (1+2)-dimensional world for the interacting particle, and the
non-minimal coupling proposed in eq. \ref{covader} selects out the electric
component of the external electromagnetic field , which allows to identify
the combination $-g\overrightarrow{v}$ as playing the role of the spin of
the test particle.

\section{\protect\bigskip Acknowledgements}

One of the authors (H. Belich) expresses his gratitude to the High Energy
Section of The Abdus Salam ICTP for the kind hospitality during the period
this work was initiated. CNPq is also acknowledged for the invaluable
financial support.

\end{document}